\documentclass[preprint2]{aastex}
\usepackage{graphicx}
\usepackage{textcomp}

\newcommand{\lwincludegraphics}[2][]{%
  \sbox{0}{\includegraphics[#1]{#2}}%
  \ifdim\wd0>\linewidth
    \resizebox{\linewidth}{!}{\box0 }%
  \else
    \leavevmode\box0
  \fi}

\shorttitle{}
\shortauthors{}

\begin{document}

\title{Faint Tidal Features in Galaxies within the
  Canada-France-Hawaii Telescope Legacy Survey Wide Fields}

\author{Adam M. Atkinson\altaffilmark{1}, Roberto G.
  Abraham\altaffilmark{1} and Annette M. N. Ferguson\altaffilmark{2}}
\altaffiltext{1}{ Department of Astronomy \& Astrophysics, University
  of Toronto, 50 St. George Street, Toronto, ON, M5S~3H4 Canada }
\altaffiltext{2}{ Institute for Astronomy, University of Edinburgh,
  Blackford Hill, Edinburgh, EH9 3HJ United Kingdom }

\begin{abstract}
  We present an analysis of the detectability of faint tidal features
  in galaxies from the wide-field component of the
  Canada-France-Hawaii Telescope Legacy Survey. Our sample consists of
  1781 luminous ($M_{r^\prime}<-19.3$ mag) galaxies in the magnitude
  range $15.5 <r^\prime <17$~ mag and in the redshift range $0.04 < z
  <0.2$. Although we have classified tidal features according to their
  morphology (e.g. streams, shells and tails), we do not attempt to
  interpret them in terms of their physical origin (e.g.  major versus
  minor merger debris).  Instead, we provide a catalog that is
  intended to provide raw material for future investigations which
  probe the nature of low surface brightness substructure around
  galaxies. We find that around 12\% of the galaxies in our sample
  show clear tidal features at the highest confidence level. This
  fraction rises to about 18\% if we include systems with convincing
  albeit weaker tidal features, and to 26\% if we include systems with
  more marginal features that may or may not be tidal in origin. These
  proportions are a strong function of rest-frame colour and of
  stellar mass.  Linear features, shells, and fans are much more
  likely to occur in massive galaxies with stellar masses
  $>10^{10.5}~{\rm M}_\odot$, and red galaxies are twice as likely to
  show tidal features than are blue galaxies.
  \end{abstract}


\section{Introduction}
\doublespace \ Colliding and interacting galaxies have been the
subject of systematic investigations by observers for over 50 years,
dating back to the the {\em Catalogue of Interacting Galaxies}
published by \citet{vv59,vv77}, and the {\em Atlas of Peculiar
  Galaxies} by \citet{arp66}.  These studies inspired pioneering
numerical simulations (e.g. \citet{tt72}) which first suggested that
elliptical galaxies may form from the coalescence of disks, thereby
changing the status of mergers from rare exotica to fundamental agents
for galaxy building.  The significance of mergers and accretions has
only grown in importance since then, as hierarchical cosmological
models have come to the forefront (e.g. \citet{whi91}). The primary
process for galaxy growth in these models is the merger and accretion
of dark matter halos. Since baryonic material traces the dark matter
(albeit imperfectly), the visible signatures of this mechanism are an
important observable useful for testing the current cosmological
paradigm.

The properties of tidal features resulting from mergers and accretions
depend on many factors such as the relative masses of the interacting
systems, the geometry of the encounter and the gravitational
potential.  In cases where mergers are of systems with roughly equal
masses, relatively high surface brightness features are expected to
form, and some of these are quite well understood. For example,
\citet{mihos98} show that for a given mass ratio the main parameter
governing the length and kinematics of tidal tails is simply the shape
of the potential, with the strength of the perturbation being
relatively independent of the encounter speed. Another class of
features that are fairly well understood are the `shell' systems seen
in elliptical galaxies (e.g. Malin \& Carter 1983). It is likely that
the azimuthal distribution of shells probes the geometry of the merger
event, so that plunging mergers result in radially organized shells,
while high angular momentum mergers result in shells distributed
quasi-randomly in azimuth (e.g. Dupraz \& Combes 1986; Hernquist \&
Quinn 1988).  Unfortunately, most mergers in the Universe are not of
equal mass galaxies, and most tidal features are not as easily
visible, nor as well understood, as tails or shells.

Cosmological simulations driven by the desire to understand
hierarchical galaxy formation naturally focus on the more common minor
mergers and accretions and the remnant substructure these leave behind
\citep{bul05,joh08,coo10}. The statistical properties of this
substructure contain an imprint of the merging history and nature of
the progenitor galaxies.  A fundamental prediction of hierarchical
galaxy formation models in a $\Lambda$-dominated Cold Dark Matter
cosmology (LCDM) is that all galaxies are surrounded by a vast and
complex network of ultra-low-surface brightness filaments and streams.
Ê A few such streams and filaments have been discovered around nearby
galaxies, including the Milky Way (e.g. \citet{bel06}) and M31 (e.g.
Ferguson et al. 2002; McConnachie et al. 2009), providing irrefutable
evidence that some systems formed, at least in part, hierarchically.
ÊHowever, other galaxies scrutinised to faint depths fail to show such
coherent debris (e.g. Barker et al. 2009, 2012; Bernard et al. 2012;
Bailin et al. 2012) and it remains unclear if the number of discrete
substructures around the galaxy population as a whole agrees with LCDM
expectations.  Bullock \& Johnston (2005) and others have pointed out
that the bulk of the substructure resulting from accretion events is
expected to have extremely low surface brightness, 30 mag/arcsec$^2$
or fainter, which is well below the limit typically reached in modern
imaging surveys.  Thus, our current understanding of the statistical
properties of very faint tidal debris is rather limited.

In between the relatively well-understood and clearly visible
structures such as tails and shells, and the yet-undetected network of
complex ultra-low-surface brightness structures predicted to exist fainter
than 30 mag/arcsec$^2$, lies the middle ground of tidal structures
that are clearly detected but whose demographics are poorly characterized. 
This middle ground represents the main focus of
this paper, whose goal is to present a catalog of such structures. 
The ubiquity and origin of features in this surface
brightness realm ($\sim 26-28$~mag/arcsec$^2$) are surprisingly poorly
understood but are of direct relevance for understanding a variety of
processes associated with the growth of galaxies. Previous attempts to
quantify the overall statistics of galaxy substructures have reported
a wide range in the fraction of galaxies which exhibit features that
are likely to be tidal in origin. Table \ref{SynthesisOfOtherWork}
presents a summary of published results.  The lack of agreement almost
certainly reflects the range in survey depths and the different galaxy
selection criteria adopted.  The widest-area surveys (such as those
based on data from the Sloan Survey) are fairly shallow and have only
modest angular resolution. Smaller studies, such as \citet{van05} and
\citet{tal09}, go deeper, sometimes probing RMS variations in the sky
down to $\sim28$ mag/arcsec$^2$ by stacking multiple bands together. 
In general, the deepest investigations focus on sub-populations, such as
early-type galaxies or those existing in dense environments, which
appear to show statistically significant differences in their fine
structure fractions relative to the total galaxy population
\citep{van05,nai10,adams12}.

\begin{deluxetable}{lcllll}
\tabletypesize{\tiny}
\tablewidth{0pt}
\tablecaption{Overview of faint substructure from published
  surveys\label{SynthesisOfOtherWork}} \tablehead{
  \colhead{Investigation} & \colhead{N} & \colhead{Dataset} &
  \colhead{Selection} & \colhead{SB limit} &
  \colhead{Summary}\\
  \colhead{} & \colhead{} & \colhead{} & \colhead{} &
  \colhead{[mag/arcsec$^2$]} & \colhead{} } 
  \startdata
This paper  	& 1781  	& CFHTLS-Wide 	&  $r'<17$ 		& $g'_{AB}\sim27.7$ 					& Strong: 17.6\% \\
		  	&   	       	&   		          	& $0.04<z<0.2$       	&$(g'r'i')_{AB}\tablenotemark{a}\sim27.3$     	& Weak: 25.2\% \\
			&		&				&M$_r'<-19.3$		&									&\\
\\
\citet{adams12}	&3551	&CFHT MENeaCS	& Cluster ETGs	 	& $r'_{AB}\sim26.5$ 						& 3\% \\
			&		&				& $0.04<z<0.15$	&									&\\
			&		&				&M$_r'<-20$		&									&\\
\\
\citet{sheen12} 	& 273       & CTIO Mosaic II       & Cluster ETGs	& $r'_{AB}\sim30$ 						&Total Sample: 25\% \\
			& 		&				&z$\lesssim0.1$	&									& Bulge-dominated: 38\% \\
			&		&				&M$_r'<-20$		&									&\\
\\
\citet{kim12} 	& 65 		& $S^4G$ 		& ETGs 			& $[3.6\mu]_{AB}\sim26.5$ 				& 17\% \\
\\
\citet{mis11}	& 474   	& SDSS DR7       	& Edge-on disks    	& $(g'r'i')_{AB}\tablenotemark{a}\sim26$\tablenotemark{b} 	& Strong: 6\% \\
			&       	&                		& $>2'$ diameter      	&                 										& Weak: 19\%\\
\\
\citet{nai10}  	& 14034 	& SDSS DR4       	& $g' < 16$;  		 &  $g'_{AB}\sim26.5$         				& 7\%\\
			&       	&                		&  $0.01<z<0.1$        &                 							&  \\
\\
\citet{bri10}    	& 23854 	& CFHTLS-Deep    	& i$_{vega}$ \textless 21.9;       & $g'_{vega}\textless29$\tablenotemark{c}    	& 4.3\% (z=0.3) \\
			&       	&                		& 0.1 \textless z \textless 1.2;   	&                 							& 19\% (z=1) \\
			&      		&                		& M$_\star > 10^{9.5}$ M$_\odot$   									&                    \\
\\
\citet{tal09}     	& 55    	& SMARTS (1 m)   	& Ellipticals       	& $V_{vega}\sim27.7$            	& 73\%\\
			&       	&                		& $M_B<-20$           	&                 				& \\
			&       	&                		& $15<D_L<50$Mpc &                 				& \\
\\
\citet{van05}&126&MUSYC +&ETGs &(\emph{BVR}$_{AB}\tablenotemark{a}$, \emph{BVI}$_{AB}\tablenotemark{a}$)$\sim 28$&Total sample: 53\% \\
		   &      & NOAO Deep-Wide & R \textless 17         										  &              & Bulge dominated: 71\% \\
		   &	   & 				   &$0.04<z<0.2$ & & \\
\\
\\
\citet{sch88}  	 & 74    & KPNO 0.9m     & E/S0                       &  IIIaJ$\sim26.5$       	& Strong: 16\% \\
		     	 &         &                         &                                &                 		       	& Weak: $>50$\%\\
\\
\citet{mal83}	& 327  & UK Schmidt     & E/S0                     	& IIIaJ$\sim26.5$           	& 5.8\% (Shells only)\\
\citet{mal83} 	& 73    & UK Schmidt     & E/S0 (isolated)       	& IIIaJ$\sim26.5$            	& 16.5\% (Shells only)\\

\enddata
\tablenotetext{a}{Stacked.}
\tablenotetext{b}{Detections decline after this surface brightness, but individual features were detected down to nearly 28 mag/arcsec$^2$.}
\tablenotetext{c}{Cosmological dimming at the mid-redshift point of their sample (z=0.65) is 2.2 mag/arcsec$^2$, considerably higher than any other entry (e.g. 0.4 mag/arcsec$^2$ for of our sample's mid-redshift point of z=0.1).}

\end{deluxetable}

The statistical characteristics of faint tidal substructures are
surprisingly poorly understood. The lack of concordance in Table~1 highlights the need for a better
understanding of the frequency and nature of tidal debris, including
how these vary as a function of galaxy stellar population, mass and
environment. This is the main focus of this paper, the goal of which
is to provide a uniform catalog of tidal structures, grouped logically
into descriptive forms.  It is tempting to seek a design language for
describing mergers based on computer-based classification techniques
(e.g. Abraham et al. 1996; Abraham, van den Bergh \& Nair 2003).
Ultimately, that would seem the be the best way forward, but at
present the state-of-the-art remains visual
classification\footnote{Assuming data volume and classification time
  are not factors.}.  The human eye-brain system has a remarkable
sensitivity to faint, unique features, many of which escape automatic
classification using presently available techniques (e.g. see the
discussion in Adams et al. 2012).  In devising categories for
describing tidal features one should seek to assign structures to
groups with physical relevance, but it is important to acknowledge
from the outset that faint tidal debris may have a variety of origins.
With the exception of tails (and possibly shells) it is not presently
possible to definitively associate most forms of tidal debris with a
particular type of interaction (e.g. major vs. minor mergers, plunging
vs tangential orbits, grazing encounters etc.) based on morphology
alone. We agree with the viewpoint espoused by Kormendy (1982), who
noted that
\begin{quote}``... morphology is more generally a `soft' science,
  which is best viewed as preparation for more quantitative work. Its
  most important use may be to provide a list of specific questions
  which provide direction for this work.''\end{quote} Nevertheless, it
seems reasonable to expect that some classification categories are
more insightful than others. For example, Tal et al. (2009) suggests
that diffuse structures surrounding early-type galaxies may mostly be
the product of gas-free (`dry') merging events, so it seems worthwhile
to try to note the existence of such features when trying to
understand the demographics of tidal structures. Therefore, in the
present paper we categorize structures in descriptive ways that we
hope will prove meaningful in future investigations. But we emphasize
that our primary goal is a descriptive census of faint debris,
regardless of origin, and defer to future work any serious attempt to
interpret these structures.

To place the data upon which the present paper is based in some
context, we note that the most comprehensive visual catalog of faint
galactic features is currently that given in \citet{nai10}, which is
based on data from the Sloan Digital Sky Survey (SDSS). The SDSS
images are fairly uniform, but also relatively shallow and their
typical image quality (median r-band seeing of 1.43 arcsec) is not up
to the standards of the best ground-based observatories. Therefore the
data in
the present paper explores
tidal features in galaxies at redshifts similar to that in
\citet{nai10}, but probing to substantially deeper surface brightness
levels using data with better image quality (median $r^\prime$-band
seeing of 0.81 arcsec).


All magnitudes used in this paper are based on the AB system unless
otherwise noted. All cosmological calculations assume a flat dark
energy-dominated cosmology with $\Omega_{\rm M}=0.3$,
$\Omega_\Lambda$=0.7 and ${H_0}=70\ {\rm km}\ {\rm
  s}\textsuperscript{-1} {\rm Mpc}\textsuperscript{-1}$.

\section{The Data}
\subsection{CFHTLS-Wide}
The Canada-France-Hawaii Telescope Legacy Survey (CFHTLS) is the
product of 450 nights of observations on CFHT from 2003-2009 using the
MegaCam one square degree visible-wavelength imager \citep{bou03}. The
survey was split into three parts: {\em Very Wide} (a shallow survey
intended to be most useful for investigations of Kuiper-belt objects),
{\em Wide} (optimized for studies of weak lensing), and {\em Deep}
(optimized for synoptic investigations of distant supernovae over a
relatively small area). In this paper we focus solely on the
intermediate-depth data from Wide component of the survey (hereafter
referred to as CFHTLS-W). This survey covers approximately 170 square
degrees of the sky in four separate patches ranging in size from 25 to
72 square degrees.  This paper focuses on CFHTLS-W instead of the deep
synoptic survey because the large area of CFHTLS-W allows us to define
a sizeable sample of nearby objects probed with sufficient spatial
resolution in the rest-frame to allow the detection of thin features.
Tidally disturbed galaxies at high-$z$ in the deep synoptic survey
have already been investigated by Bridge, Carlberg \& Sullivan (2010),
and in \S 4 of this paper we will compare our low-redshift results to
their high-redshift results.

\begin{figure*}[!htb]
\lwincludegraphics{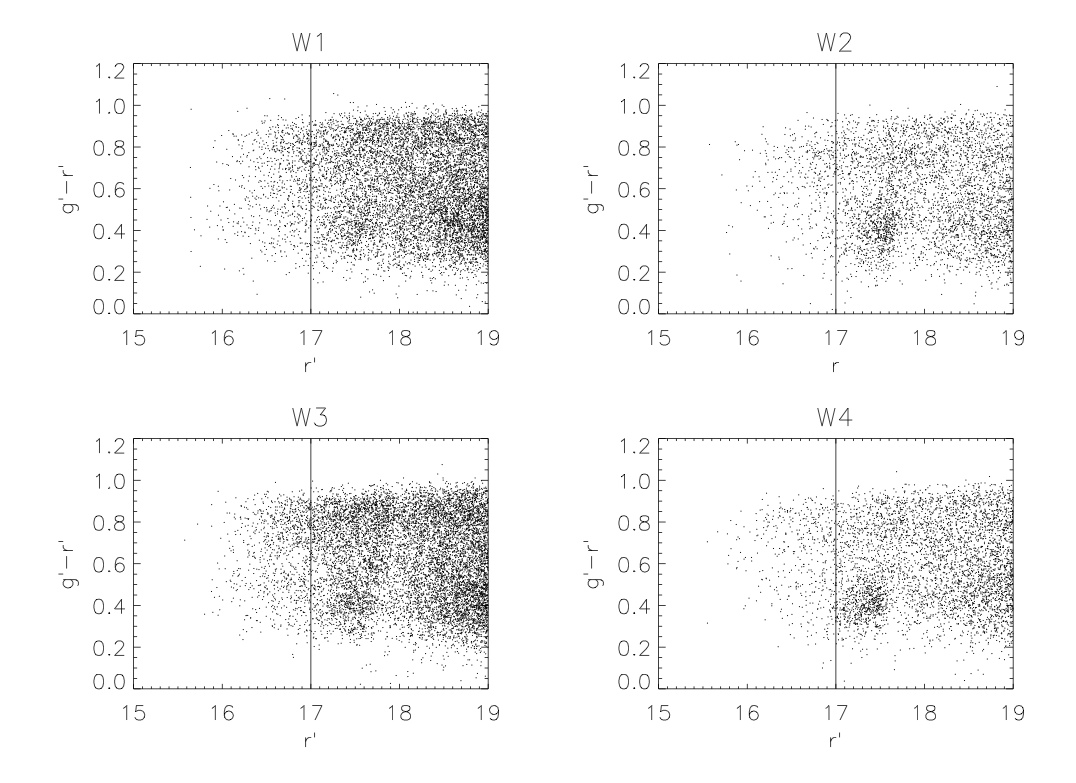}
\caption{Colour-magnitude plots of galaxies in the four CFHTLS wide fields selected 
to be between the redshifts of 0.04 and 0.2. We define our sample as galaxies with 
an $15.5 < r^\prime<17$ mag. No initial cuts were made based on colour or morphology. 
Our sample is defined by points left of the vertical line at $r^\prime$=17 in each graph. 
After unusable thumbnail images are removed we are left with a sample of 1781 galaxies 
to be visually examined for streams. The cloud of points located from $17< r^\prime<18$ 
mag and $0.2 < (g^\prime - r^\prime ) <0.4$ appear to be stars that were incorrectly labelled 
as galaxies in the photo-z catalog. No automated process was found for removing such stars 
in our sample so they were removed manually during the inspection process.
\label{CM}}
\end{figure*}

The CFHTLS-W survey is based on data obtained through five filters
($u^*,g^\prime,r^\prime,i^\prime,z^\prime$) with the Wide survey total
exposure time being approximately one hour per filter per field. The
four wide fields were centered at the following J2000 coordinates: RA
02:18:00, Dec -07:00:00 (W1); RA 08:54:00, Dec -04:15:00 (W2); RA
14:17:54, Dec +54:30:31 (W3); and RA 22:13:18, Dec +01:19:00 (W4). Our
analysis is based on publicly-available image stacks provided by the
MegaPipe pipeline
\citep{gwy08}\footnote{http://www3.cadc-ccda.hia-iha.nrc-cnrc.gc.ca/community/CFHTLS-SG/docs/cfhtls.html},
to whom the reader is referred for details on the data reduction.
Spectroscopic redshifts were unavailable for most galaxies in the
sample, so publicly-available photometric redshifts provided by
\citet{cou09} were used instead. When needed, K-corrections were
applied to the data using the program detailed in \citet{chi10}.

\subsection{Sample Selection}
\singlespace

\begin{figure}[!htb]
\lwincludegraphics{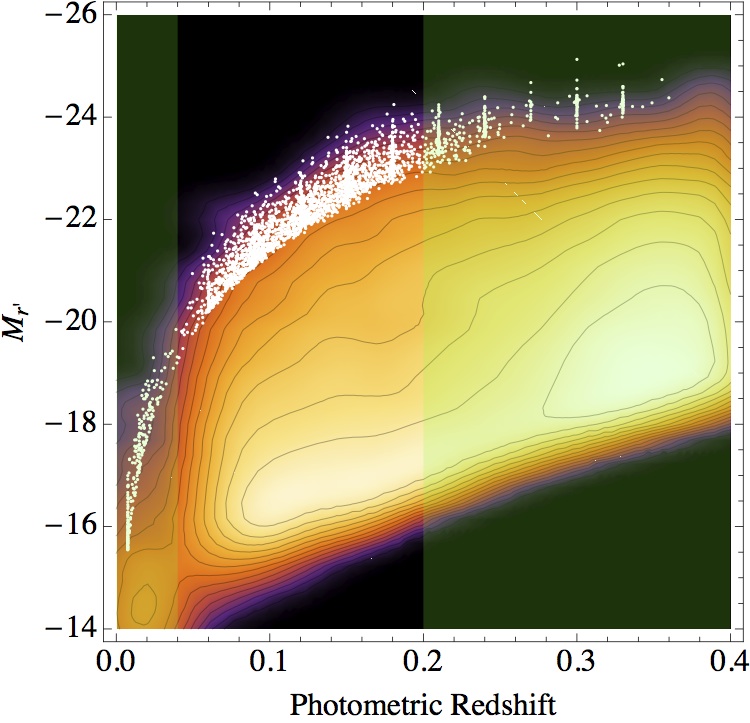}
\caption{Absolute magnitude as a function of photometric redshift for galaxies which are brighter than the apparent magnitude limit of our 
sample (white points) superposed on the distribution for the full CFHT Legacy Survey sample out to $z=0.4$, shown as a smoothed density 
histogram (with logarithmically-spaced contours). Redshift cuts ($z<0.04$ and $z>0.2$) used to excise
very nearby and distant galaxies are shown as translucent green bands. Note how the final sample probes mainly the bright end of the galaxy 
distribution. See text for details.
\label{density}}
\end{figure}

\doublespace Objects were selected for inspection by filtering the
photometric redshift catalog to restrict consideration to objects in
the redshift range $0.04< z<0.2$.  This range overlaps with that of
Nair \& Abraham (2010) and allows for a direct comparison with
\citet{van05}. An upper magnitude limit of $r^\prime=17$ mag was
imposed in order to limit the sample size to a manageable number for
visual inspection. A lower magnitude cutoff was imposed at
$r^\prime=15.5$ mag because below this limit stars misidentified as
galaxies begin to outnumber galaxies in the photometric redshift
catalog. The implications of these cuts are illustrated in Figure
\ref{CM}, which shows a color-magnitude diagram with included objects
lying to the left of the vertical cut in each panel. (Note that
objects failing to meet the photometric redshift or the bright
magnitude cuts have already been removed in this figure). After
removal of image artifacts (e.g. ghost images from bright stars,
segments of diffraction spikes), misidentified stars, and otherwise
unusable images (e.g., fields that are too crowded with foreground
stars) 1781 galaxies remained in our sample.  Our sample is heavily
biased toward bright systems, as shown in Figure~\ref{density}. This
diagram compares the absolute magnitude as a function of redshift for
our sample against the corresponding distribution for the full CFHT
Legacy Survey. Most of the sample lies in the range
$-23<M_{r^\prime}<-20$ mag.  Our low-redshift cut eliminates most
under-luminous galaxies, so the faintest galaxy in our sample has an
absolute magnitude of $M_{r^\prime}=-19.3$ mag\footnote{Note that
  Figure~\ref{density} clearly shows evidence for some discreteness in
  the photometric redshifts in the \citet{cou09} catalog, but this is
  at a level that is insignificant for our purposes.}. On the other
hand, the absolute magnitudes of objects in our sample of typical of
those of brighter galaxies seen at high redshifts, so our dataset is
well-suited for comparisons made against these.  The typical
half-light radii for our galaxies is 2--6 arcsec.

\begin{figure}[!htb]
\plotone{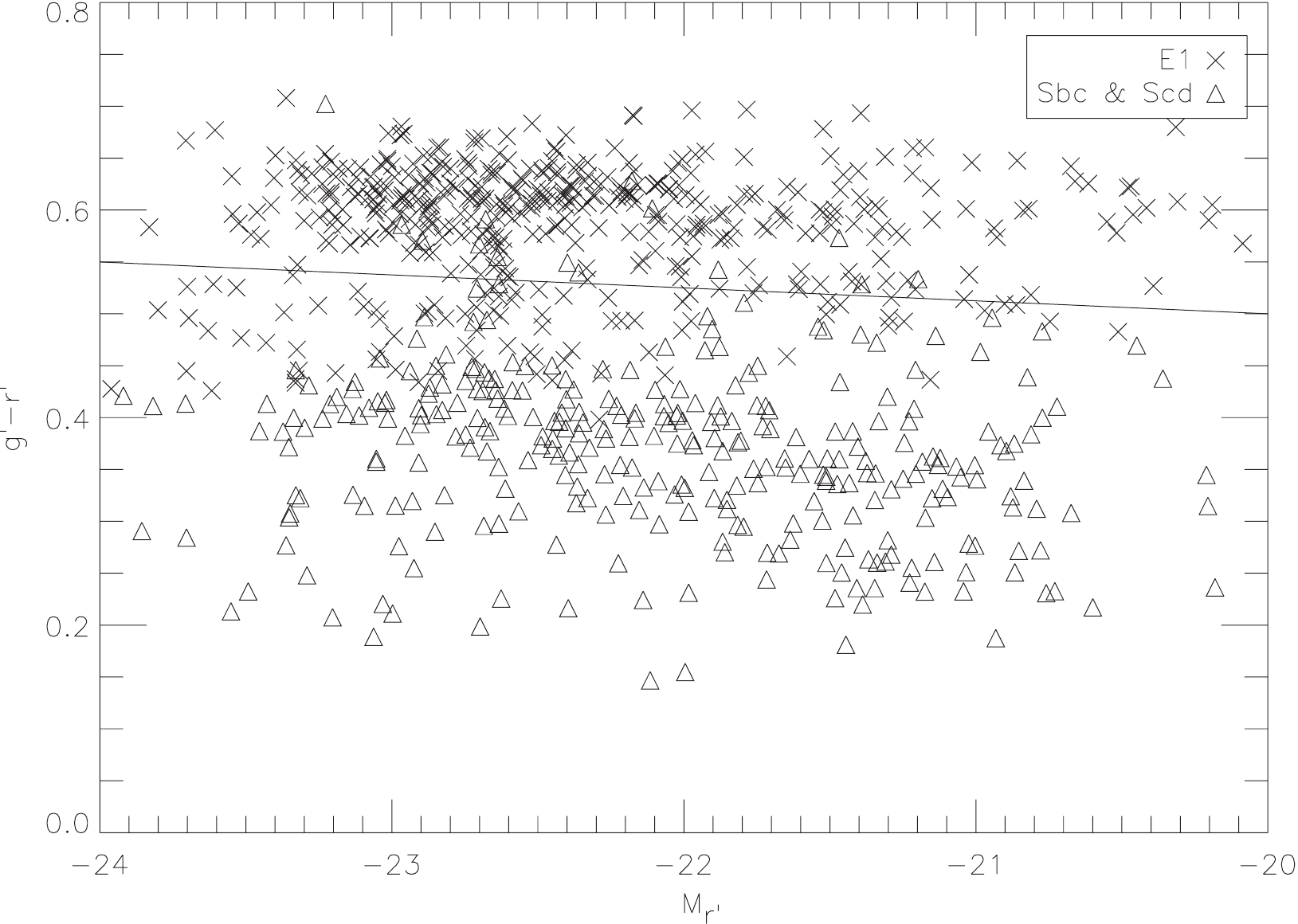}
\caption{Colour-magnitude diagram for galaxies in the CFHTLS-W1 field. The line shown corresponds to $(g'-r')= -0.0125 r' + 0.25$. 
As described in the text, this line is used to divide the total sample into red and blue sub-populations, corresponding to systems on the 
`red sequence' and the `blue cloud'. Plot symbols are keyed to galaxy templates used to determine photometric redshifts by Coupon et al. (2009).
Galaxies identified by these authors as being best fit by the E1 template are plotted with crosses, while those best fit by Sbc and Scd templates 
are plotted with triangles. The corresponding plots for the three other CFHTLS wide fields (not shown) closely resemble this one. Over the entire 
sample, approximately 86\% of red sequence galaxies were best fit by an E1 template while 84\% of blue cloud galaxies were best fit by an Sbc or Scd template. 
\label{CCL}}
\end{figure}

As will be shown below, the statistical properties of the
tidally-disturbed galaxy sample are quite strongly dependent on
rest-frame colour. It is therefore useful to subdivide the sample into
red and blue sub-populations from the outset.  Sub-populations are
defined based on position in the $(g^\prime-r^\prime)$ vs. $r^\prime$
colour-magnitude diagram, using the following line to subdivide the
galaxy population:

\begin{equation}
(g^\prime-r^\prime)= -0.0125 r^\prime + 0.25
\end{equation}

\noindent This line was defined by us to to discriminate between
systems on the red sequence and blue cloud \citep{bel04,str01}.  As
shown in Figure~\ref{CCL}, this line does an excellent job of
subdividing galaxies based on fits to the galaxy templates used by
\citet{cou09} to determine the photometric redshifts for the sample.
Because of the correlations between rest-frame color and Hubble stage,
our color-based subdivision of the galaxy population results in a
coarse morphological segregation as well. For example, 86\% of all red
sequence galaxies were best fit with an E1 template by \citet{cou09},
while 84\% of all blue sequence galaxies were best fit by an Sbc or
Scd template.

\subsection{Visual Inspection}

\begin{figure}[!htb]
\plotone{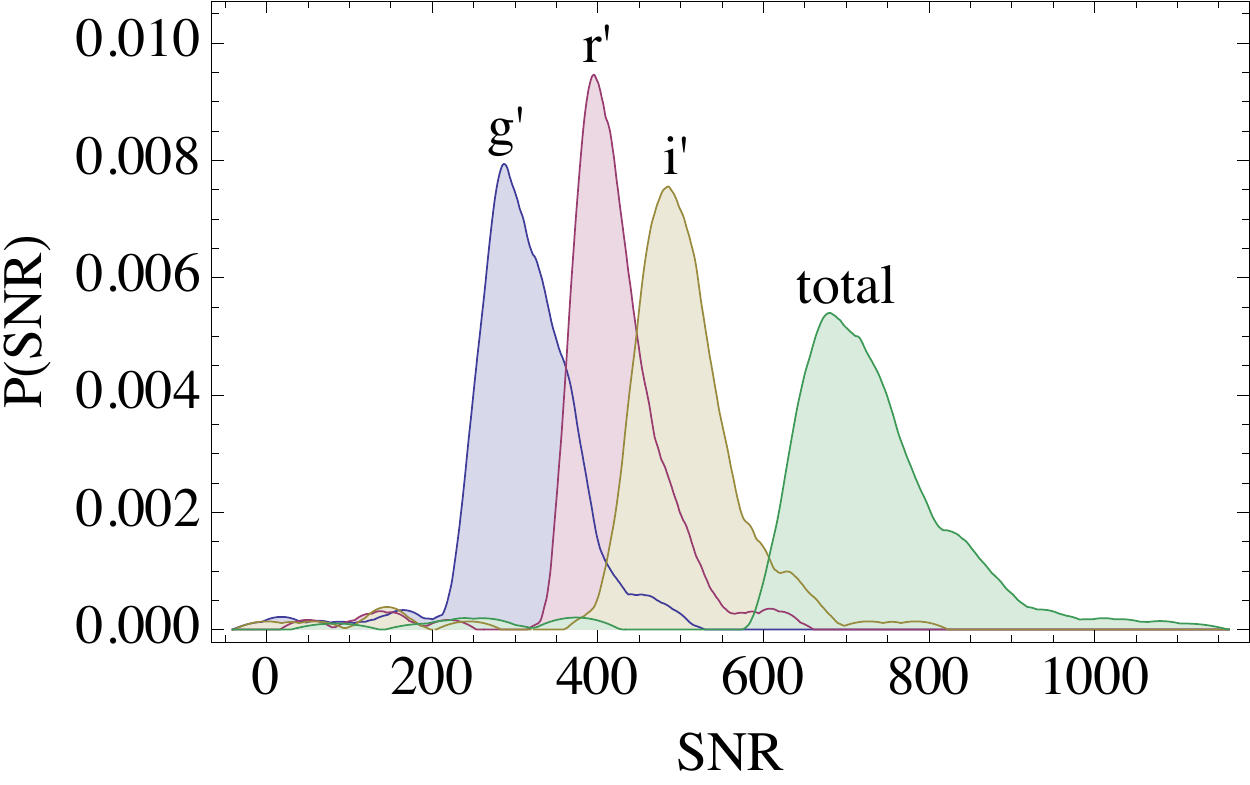}
\caption{Smoothed probability density functions showing the signal-to-noise ratios in the $g'$, $r'$ and $i'$-band 
images and the co-added $g'+r'+i'$ stack. Histogram smoothing is done using an Epanechnikov kernel density estimator. 
See text for details. \label{snr}}
\end{figure}

Thumbnail images in the $g^\prime$, $r^\prime$ and $i^\prime$ band of
each galaxy were cut out of the CHFTLS-Wide fields and stacked
together (i.e. summed) to increase contrast. The relatively low
signal-to-noise $u^*$ and $z^\prime$ images were not co-added, as the
gain in signal-to-noise obtained was deemed insufficient. For each
thumbnail the signal-to-noise level was estimated by using {\tt
  SExtractor} to compute the ratio of the isophotal flux to the error
on the isophotal flux reported by the program\footnote{In other words,
  {\tt FLUX\_ISO/FLUXERR\_ISO}.} \citep{ber96}. Distributions of SNR
for the thumbnails are shown in Figure~\ref{snr}. As expected, the
total signal to noise of the galaxy in the stacked frame is
well-approximated by simply adding the individual signal-to-noises of
the filtered data in quadrature. The median signal-to-noise ratio of a
stacked galaxy in our sample is 712, which is a remarkably close match
to the mean signal-to-noise ratio of galaxies in the Nair \& Abraham
(2010) catalog of morphological classifications from the SDSS ($712$
for $g'$-band, 735 for $r'$ band, 592 for $i'$-band). Note that Nair
\& Abraham (2010) did not stack images in different filter bands to
increase signal-to-noise.

\begin{figure*}[!htb]
\lwincludegraphics{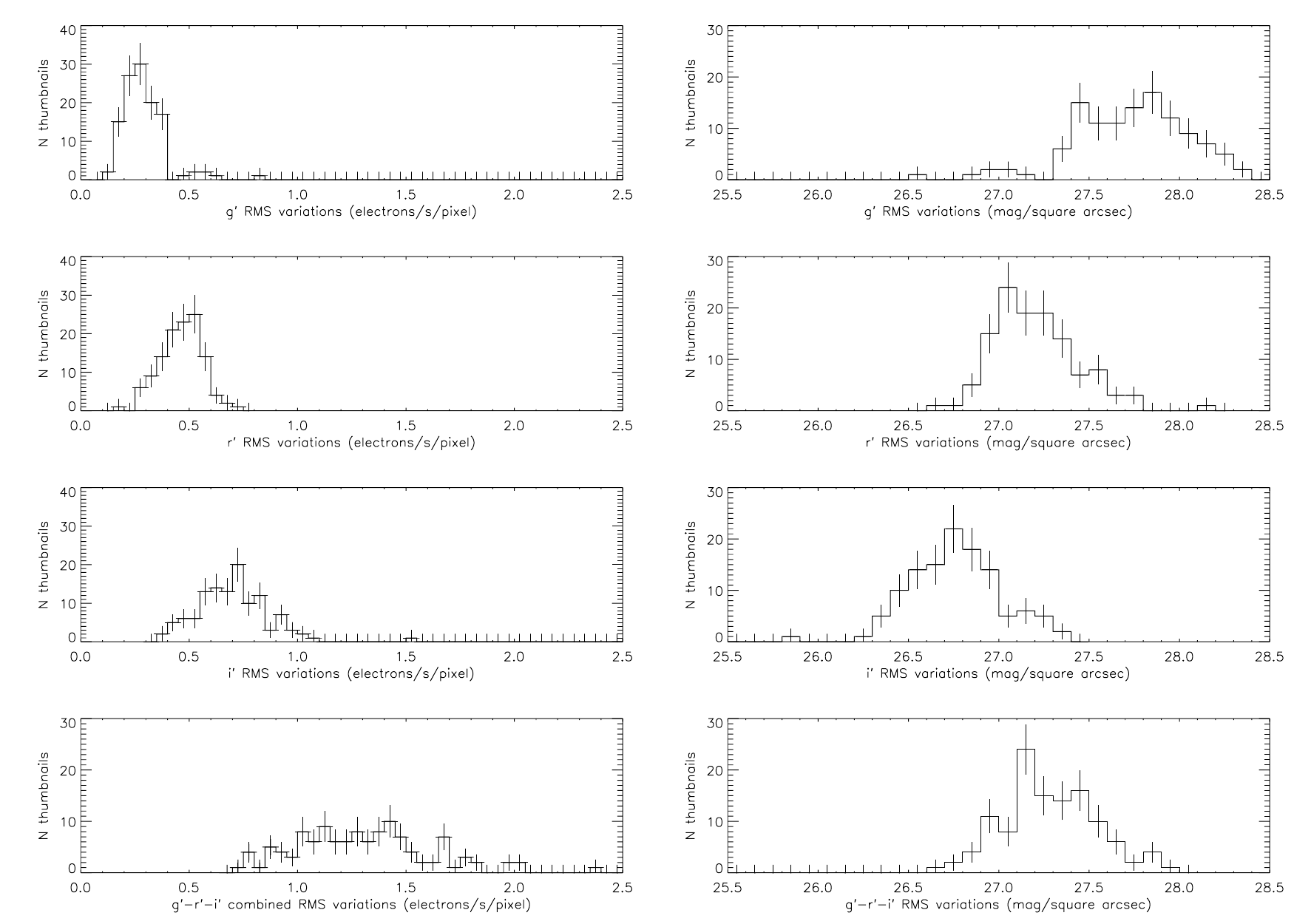}
\caption{Histogram of the RMS variations due to sky noise estimated
  from 120 thumbnail images of separate galaxies representing all four
  wide fields. Noise for each of the thumbnails was found by placing
  40 1.2 square arcsecond apertures in each image, and then computing
  the standard deviation of the mean. The left-hand column presents
  data in electron/s/pixel, while the right-hand column shows the
  corresponding values in mag/arcsec$^2$. Since the $g'$ + $r'$ + $i'$
  stack does not correspond to a standard photometric band the
  definition of an AB magnitude is used to backtrack an equivalent
  zero point. Rows correspond to $g'$, $r'$, $i'$, and stacked
  $g'$+$r'$+$i'$. See text for details. \label{hists}}
\end{figure*}

To determine the limiting depth of the data, 40 square apertures, each
with an area of $1.2$ square arcsec were placed in empty regions of
120 thumbnails chosen to equally represent all four wide fields.  The
RMS variations of the total enclosed flux within these apertures, for
each thumbnail image, were then computed to serve as a representation
of the sky noise for each thumbnail image. Results are shown in Figure
\ref{hists}.  In the $g^\prime$ band we find a mean limiting surface
brightness of $\sim27.7$ mag/acsec$^2$ (which corresponds to
$\sim0.5\%$ of the g-band night sky brightness) with a standard
deviation of $\sim 0.5$ mag/sq arcsec. This provides a good
representation of the limiting depth to which we can detect features
in this band on small scales, although in practise large features can
be detected to even lower surface brightness limits. Note that the
limiting surface brightness of the $g'$ data is fainter than the
limiting surface brightness of the data obtained in redder bands
(Figure \ref{hists}), but the signal-to-noise (Figure \ref{snr}) level
of the redder bands is higher.  This is simply due to the zero points
underlying astronomical magnitude systems, which are biased toward
blue objects (Vega in the traditional system, or flat spectrum sources
in $F_\nu$ in the AB system).

The thumbnails were visually inspected in the viewing program DS9
\citep{joy03} to allow interactive manipulation of the screen transfer
function, and the images were grouped into one of five bins
corresponding to the inspector's confidence that a tidal feature was
present. The five confidence bins are defined in Table
\ref{ConfidenceIntervalDefinitions}. Confidence bin 4 corresponds to
the inspector being certain that the galaxy in question contains a
tidal feature. Confidence level 3 corresponds to a galaxy that is very
likely contains a feature, with something around 75\% certainty. In
our opinion it is reasonable to group confidence bins three and four
together to define a sample with strong indications of tidal features.
Confidence level 2 indicates a 50\% likelihood that a tidal feature is
present. Treatment of such systems is naturally problematic, and it is
left up to the reader to determine whether these objects should
considered to be detections or non-detections. In the present paper we
treat them as non-detections. Confidence level 1 corresponds to there
being a hint that a tidal feature is present, while level 0 indicates
no visible tidal features (or, alternatively, strong confidence that
no features are present down to the surface brightness levels probed
by our data).  Studies of the reliability of morphological
classifications (e.g. Naim et al. 1995) show that the internal
consistency of a well-trained individual's morphological
classifications tends to be higher than the observer-observer
consistency of morphological classifications made by similarly
well-trained people working on the same data. Therefore it is often
unwise to have a few individuals visually classify the same sample and
then `vote' to determine a final classification. However, in the
present instance, there is no well-defined classification scheme (such
as the Hubble system) in place which describes tidal features. Under
the circumstances, it was felt that some preliminary inter-comparison
of individual classifications by separate individuals was in order.
Therefore 138 galaxies were examined by all three authors in order to
verify the reproducibility of the most basic classification attempted,
which is simply the observer's confidence that the galaxy being
inspected shows evidence for any form of tidal disturbance.
Excellent agreement was found on systems classified as 3 or 4 (high
confidence that tidal features exist) and on systems classed as 0
(high confidence that no features exist). As might be anticipated, the
only significant disagreements occurs for systems classified as 1 and
2 (objects with possible hints of tidal features). Encouraged by these
comparisons, all 1781 galaxies were then inspected by a single
classifier (Atkinson) in order to ensure consistency.

\begin{figure}[htb]
\begin{center}
\includegraphics[height=2.2in]{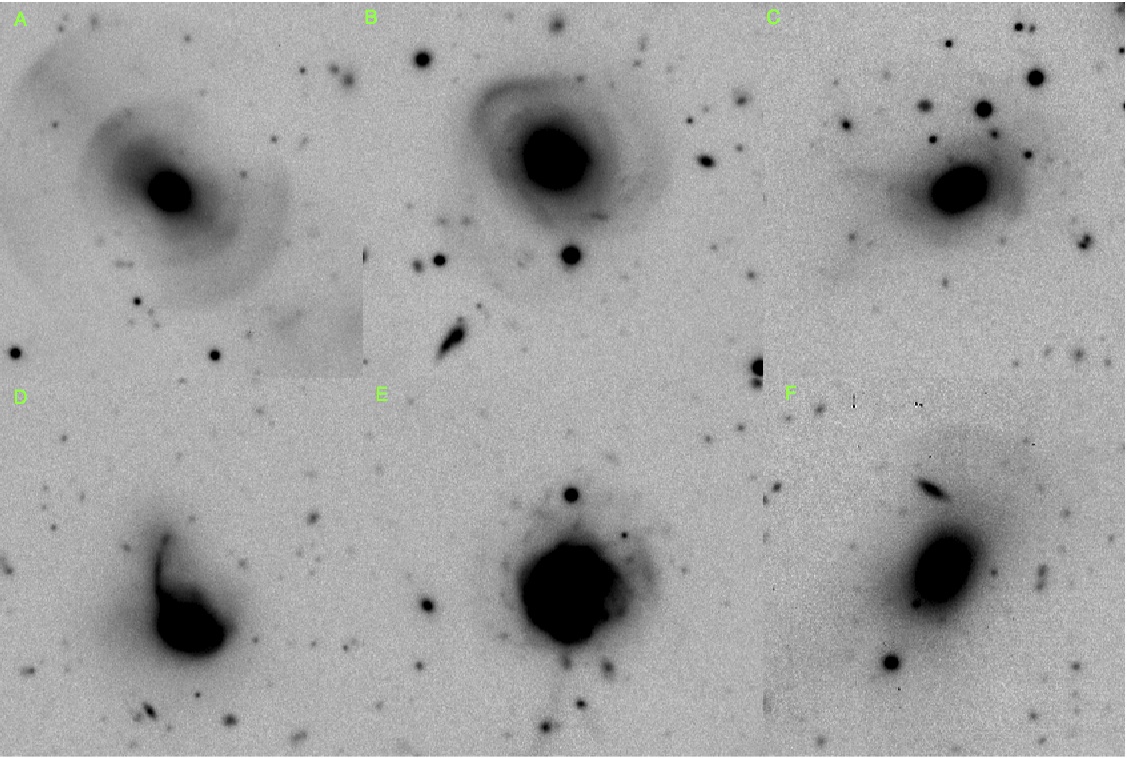}
\caption{Examples of the six different categories of tidal
  disturbances used to classify objects in this paper. (a) Shells
  surrounding a galaxy, (b) A stream (visible in the lower part of the
  image), (c) Miscellaneous diffuse structure, (d) An arm, (e) A
  linear feature (visible at the bottom of the image), (f) Broad fans
  of diffuse light. Each of the six thumbnails is approximately 56x56
  arcsec. \label{feat}}
\end{center}
\end{figure}

\begin{figure}[!htb]
\begin{center}
\includegraphics[height=5.0in]{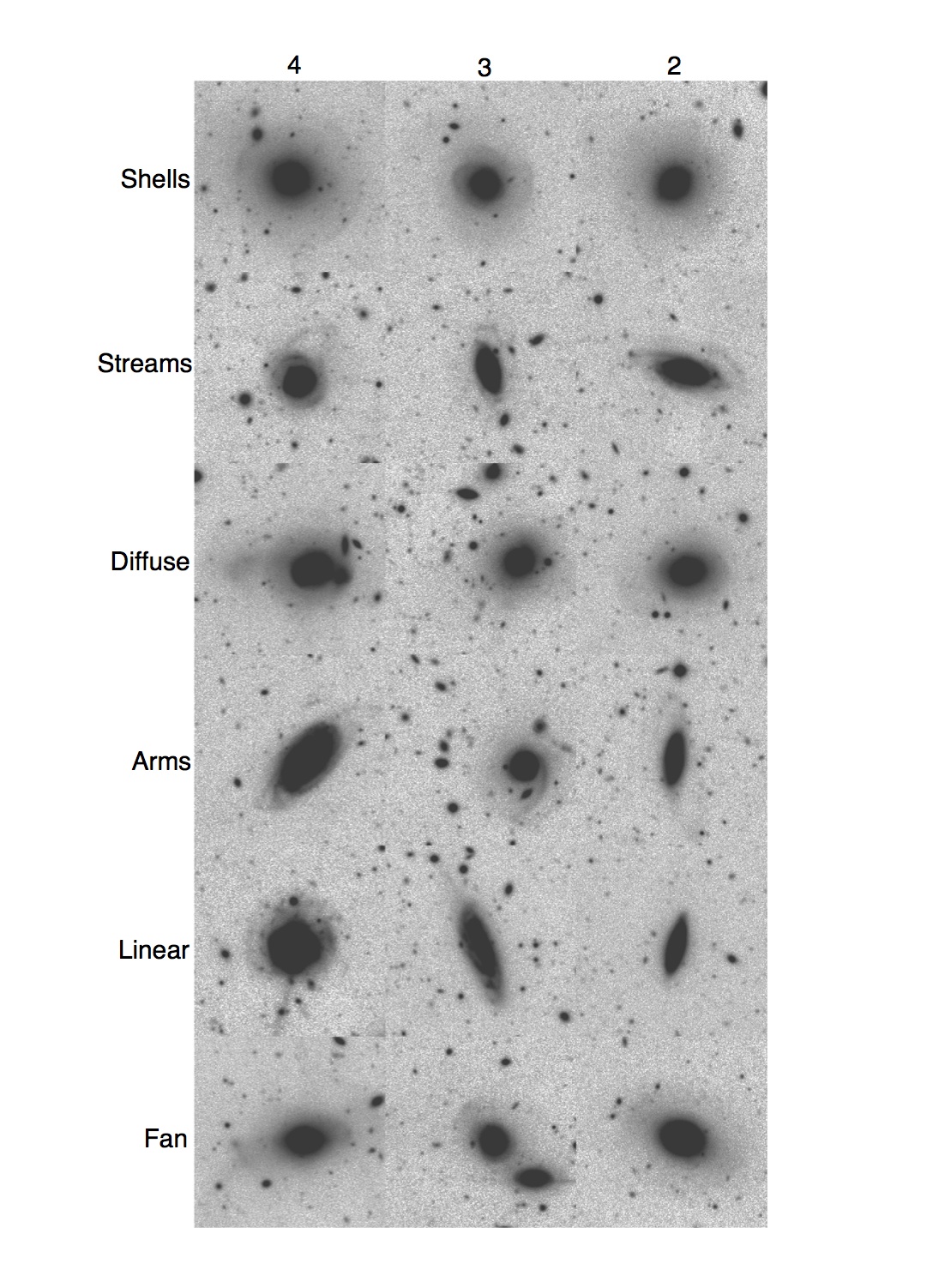}
 \caption{Examples of all six classifications of tidal features (top
   to bottom: shells, streams, miscellaneous diffuse structure, arms,
   linear features and fans) in three different confidence bins with
   decreasing confidence from left to right (bins four, three and two.
   Columns correspond to the confidences levels (defined in Table
   \ref{ConfidenceIntervalDefinitions}) that the tidal feature exists.
   Each individual thumbnail is approximately 56x56 arcsec.
\label{con}}
\end{center}
\end{figure}

\begin{deluxetable}{cl}
\tablecaption{Detection classes \label{ConfidenceIntervalDefinitions}}
\tablehead{
\colhead{Confidence level} & \colhead{Definition}}
\startdata
4 & Certain detection of a tidal feature \\
3 & Probable detection of a tidal feature (over 75 percent certain)\\
2 & Possible detection of a tidal feature (around 50 percent certain)\\
1 & Hint of a potential tidal feature. Very uncertain. \\
0 & No evidence for tidal features seen\\
\enddata
\end{deluxetable}

Tidal features were classified broadly into six categories: (1)
streams; (2) arms; (3) linear features; (4) diffuse fans; (5) shells;
and (6) miscellaneous diffuse structure. Stacked thumbnail images of
all six types of features are shown in Figure \ref{feat}, while
examples of the six features at varying confidence levels are depicted
in Figure \ref{con}.  These features are not mutually exclusive. In
fact, they should be viewed as a basis set of features from which
descriptions of more elaborate tidal structures can be composed. For
example, an `umbrella-like' structure is seen in a number of galaxies,
and this can be described as a combination of a `linear feature' and a
`shell' (a boolean `AND' operation can be used on the electronic
caltalog to do this efficiently). In other cases features can be
combined to yield broader more familiar classifications using a
boolean `OR'; for example, a set of galaxies with `tails' can be
constructed by extracting all systems with either `streams' or `arms'
or `linear features'.  While it is our hope that our basis features
might be indicative of different types of interactions (e.g., broad
fans may result from dry mergers, as claimed by van Dokkum 2005), it
is beyond the scope of our paper to test this. As noted earlier, our
aim here is to provide a basic catalog which upon which future
investigations of such questions can be based.

\section{Results}

\begin{figure}[!htb]
\plotone{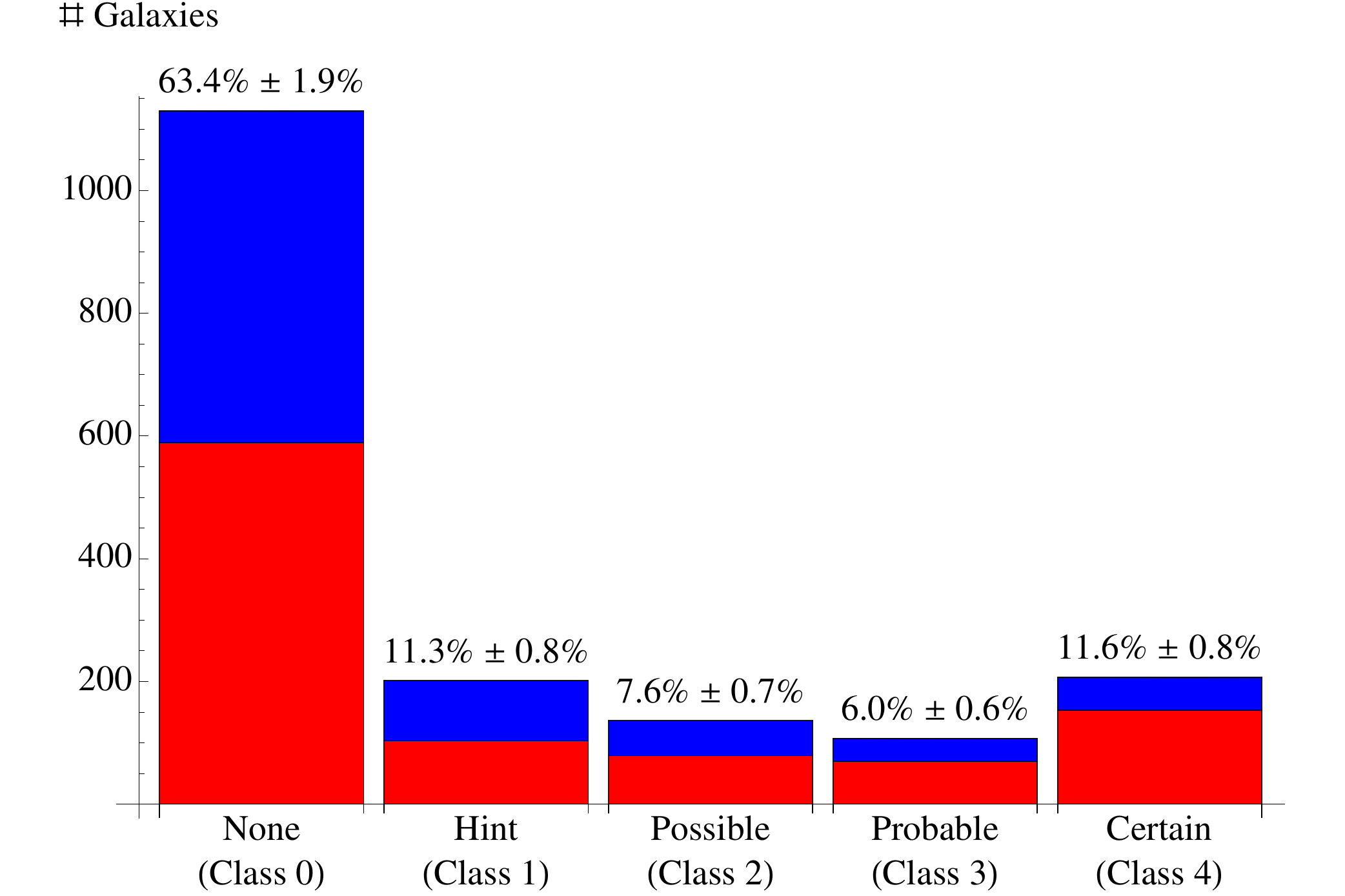}
\caption{Distribution of detection classes in the CFHTLS-Wide dataset. Each histogram bin is
labeled by its fractional contribution to the total galaxy population, and each bin
is subdivided into red sequence and blue cloud populations. 
\label{detectionHistogram}}
\end{figure}

\begin{figure}[!htb]
\plotone{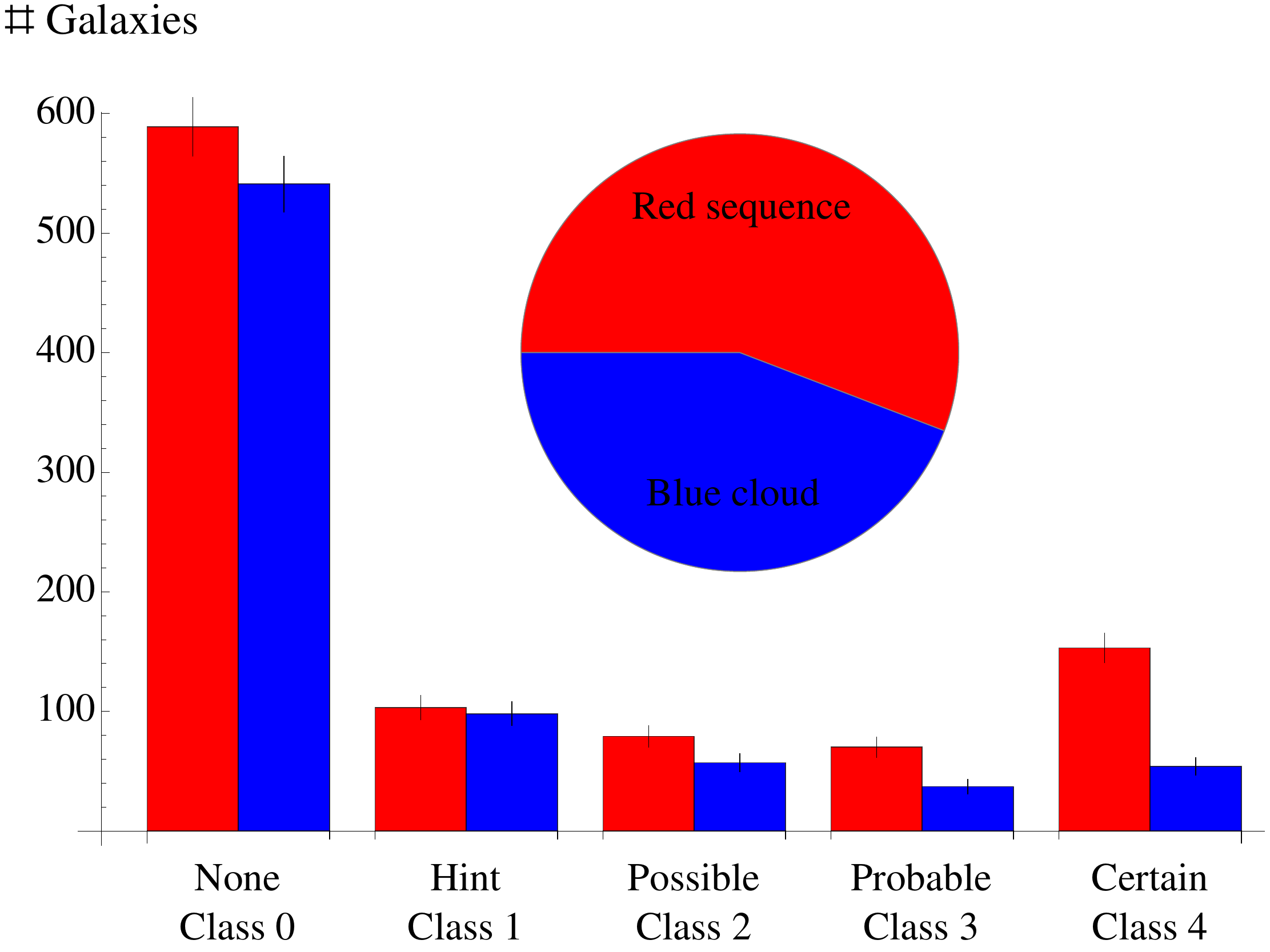}
\caption{Distribution of feature classifications. Each class of tidal feature is subdivided into
separate bins of galaxy colour.
 \label{CC}}
\end{figure}

\begin{figure}[!htb]
\plotone{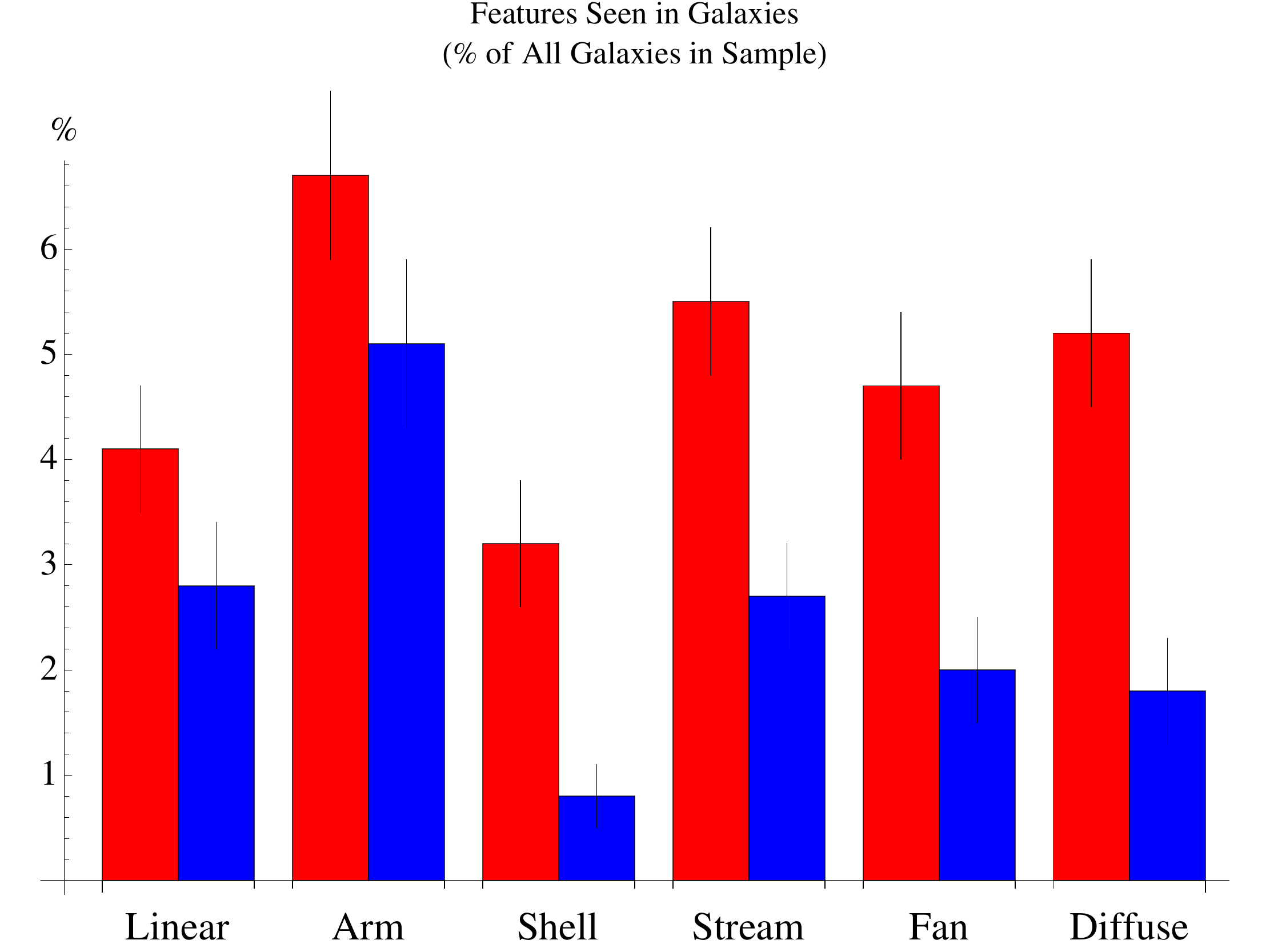}
\caption{Distribution of high-confidence (levels 3 and 4) detections of tidal features subdivided by galaxy colour. Red bars correspond to galaxies in the red sequence, while blue bars correspond to galaxies in the blue cloud. Percentages shown correspond to fractions relative to the total number of galaxies of that colour. 
  \label{CC1}}
\end{figure}

\begin{figure}[!htb]
\plotone{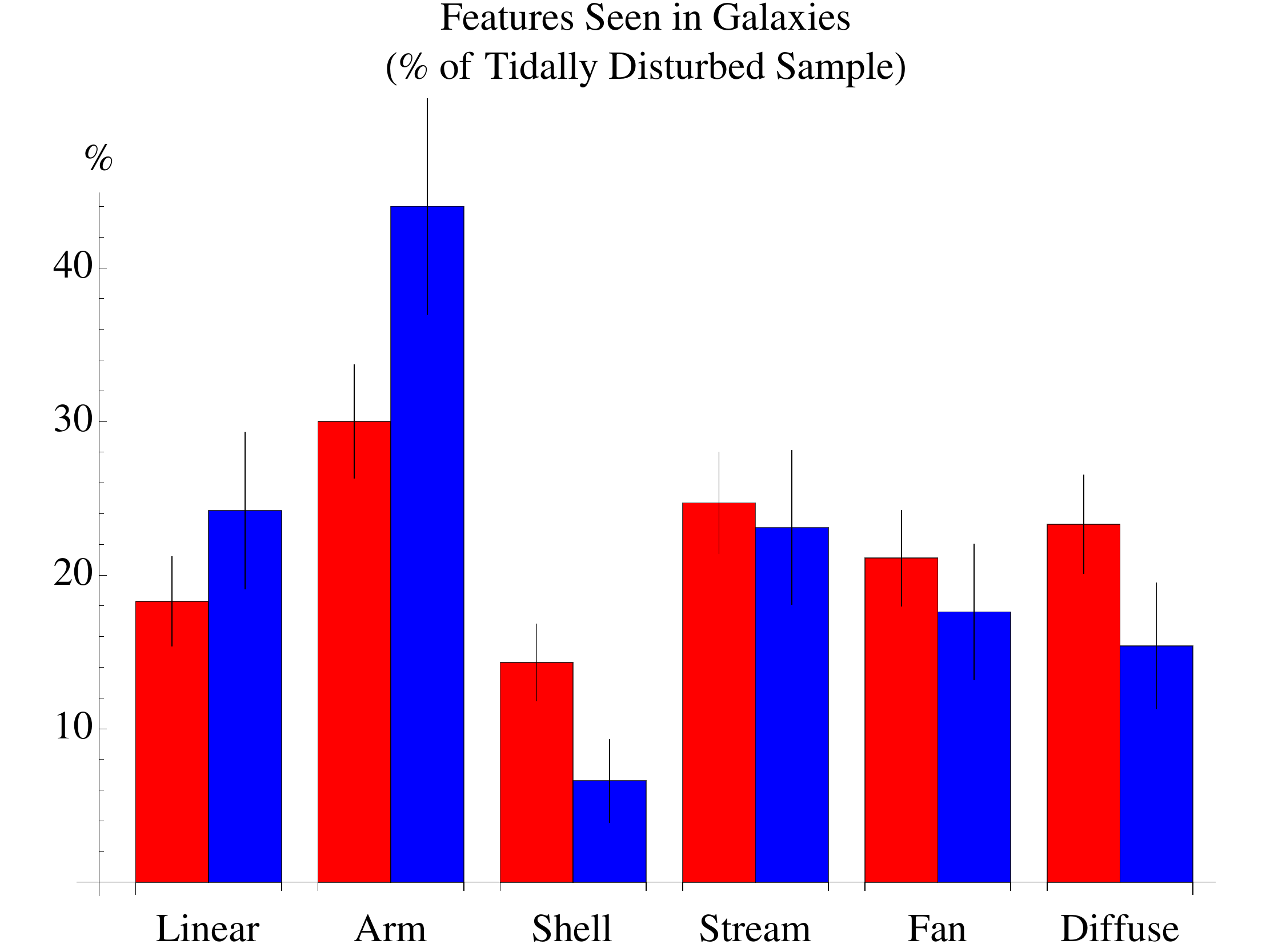}
\caption{Distribution of tidal features subdivided by galaxy colour. Red histogram bars correspond to red galaxies, while blue histogram bars correspond to blue galaxies. Percentages shown correspond to fractions relative to the number of high-confidence tidally disturbed galaxies (class 3 and 4) of that colour. 
 \label{CC2}}
\end{figure}

The central results of this paper are summarized in Table
\ref{summarystats} and shown in
Figures~\ref{detectionHistogram}--\ref{CC2}. Each of these figures is
presented in a way that allows the reader to compare the fraction of
systems showing tidal disturbances as a function of rest-frame colour.
Table \ref{MasterDataTable} lists the classifications and ancillary
data for all galaxies in our sample\footnote{An extract from this
  table is given in the print version of this paper. The full version
  appears in the electronic version.}.


\begin{deluxetable}{ccccccccc} 
\tablecolumns{9} 
\tablewidth{0pc} 
\tablecaption{Summary of Tidal Feature Detections\label{summarystats}} 
\tablehead{ 
\colhead{}    &  \multicolumn{2}{c}{Total Sample}    & \colhead{} & 
\multicolumn{2}{c}{Red Galaxies}  & \colhead{} & \multicolumn{2}{c}{Blue Galaxies}\\ 
\cline{2-3} \cline{5-6} \cline{8-9} \\ 
\colhead{Confidence} & \colhead{Number}   & \colhead{\%}    & \colhead{} & \colhead{Number} & 
\colhead{\%}    & \colhead{} & \colhead{Number}   & \colhead{\%}}
\startdata 
4 & 207\phn  &  $11.6\pm0.8$& &153 & $15.4\pm1.2$  & & 54&  $6.9\pm0.9$\\
3 & 107\phn  & $6.0\pm0.6$ &  &70  & $7.0\pm0.8$     &&  37&  $4.7\pm0.8$\\
2 & 136\phn  & $7.6\pm0.7$ &  &79  & $7.9\pm0.9$     &&  57&  $7.2\pm1.0$\\
1 & 201\phn  & $11.3\pm0.8$ & &103 & $10.4\pm1.0$   &&  98&  $12.4\pm1.3$\\
0 & 1130\phn & $63.4\pm1.9$ & &589 & $59.3\pm2.4$   &&  541&  $68.7\pm3.0$\\
\hline
\# Galaxies & \multicolumn{2}{c}{1781} & &  \multicolumn{2}{c}{994} &  & \multicolumn{2}{c}{787}  \\
\enddata 
\end{deluxetable} 


We find that $11.6\pm0.8\%$ of galaxies in our sample show tidal
features at the highest confidence level (confidence level 4). This
fraction rises to $17.6\pm1.0\%$ if systems with with confidence level
three (objects with `probable' features) are included, and to
$25.2\pm1.2\%$ if systems with confidence level two (`possible'
features) are also included. In the Discussion section of this paper
we will consider why these fractions are well below the 50-75\% of
systems reported to have tidal features by \citet{van05} and
\citet{tal09}.

Figure~\ref{detectionHistogram} depicts our overall findings, while
Figure~\ref{CC} emphasizes that chromatic effects play a role in
defining the probability that a given galaxy will show tidal
disturbances.  Interestingly, we can now also show that the nature of
the tidal disturbances is a strong function of colour. Red galaxies
are over twice as likely to show signs of tidal disturbance compared
to blue galaxies. This effect is most striking in the high-confidence
bin (confidence levels three and four), becoming somewhat diluted in
the low confidence bins, presumably as real tidal structures become
diluted by spurious false detections.

Figures \ref{CC1} and \ref{CC2} illustrate the incidence of tidal
disturbances as a function of feature class. Figure \ref{CC1} shows
the distribution of features seen with high confidence (class 3 and 4)
relative to the total galaxy sample, while Figure~\ref{CC2} shows the
distribution of features relative to the population of tidally
disturbed galaxies of each particular color\footnote{Note that tidal
  features are not mutually exclusive, so summing the values in
  Figure~\ref{CC2} exceeds 100\%.}. These two figures are clearly
closely related, but they make different points, so we will describe
each of them in turn.

Figure~\ref{CC1} shows that every type of tidal disturbance is more
likely to be found in a red galaxy than in a blue galaxy. In some ways
this seems unsurprising -- some features, such as shells, are
generally associated with elliptical galaxies, and our analysis
confirms that shells are far more likely to appear in red galaxies
than in blue galaxies (unsurprising, since a major merger leading to
an elliptical would likely destroy a disk; c.f. Hernquist \& Spergel
1992). However, other post-merger debris features, such as bridges
(generally classified in the present paper as 'arms' or 'diffuse
structures' which connects two galaxies) and tails (generally grouped
into the `arm' and `linear' classes), are generally associated with
disks. For example, all the systems modelled in the classic paper by
Toomre \& Toomre (1972) are blue star-forming objects\footnote{Arp 295
  , M51 + NGC51995 (the `whirlpool'), NGC4676 (the `mice') and
  NGC4038/9 (the `antennae').}. Our analysis shows that these sorts of
features are also more common in red galaxies than in blue galaxies.

Our analysis indicates that {\em all} types of disturbances are more
common in red galaxies than in blue galaxies, but that red and blue
populations are markedly different in the {\em diversity} of their
tidal features. Red galaxies show a plethora of structures, with
roughly the same fraction of red galaxies exhibiting `arms',
`streams', `fans' and `miscellaneous diffuse structures'. Amongst red
galaxies, the fraction of galaxies exhibiting each of these classes is
statistically indistinguishable at the 90\% confidence level. Blue
galaxies, on the other hand, are much more likely to exhibit `arms'
than any other class of structure.

This point can be seen somewhat more clearly by inspection of
Figure~\ref{CC2}, which shows our data normalized to the number of
tidally disturbed galaxies of each colour. We see from this figure
that, in both red and blue populations, `arms' are the most common
tidal feature and `shells' are the least common tidal feature. Amongst
disturbed blue galaxies, 45\% show tidal arms, while shells are only
seen in 6\% of this population. The 7:1 difference in the occurrence
rate for these features amongst blue galaxies is in sharp contrast to
the barely 2:1 difference for these features amongst red galaxies.
Evidently tidally-disturbed red galaxies tend to exhibit a richer
variety of dynamical structures than do blue galaxies.

\section{Discussion}

\begin{figure*}[!htb]
\lwincludegraphics{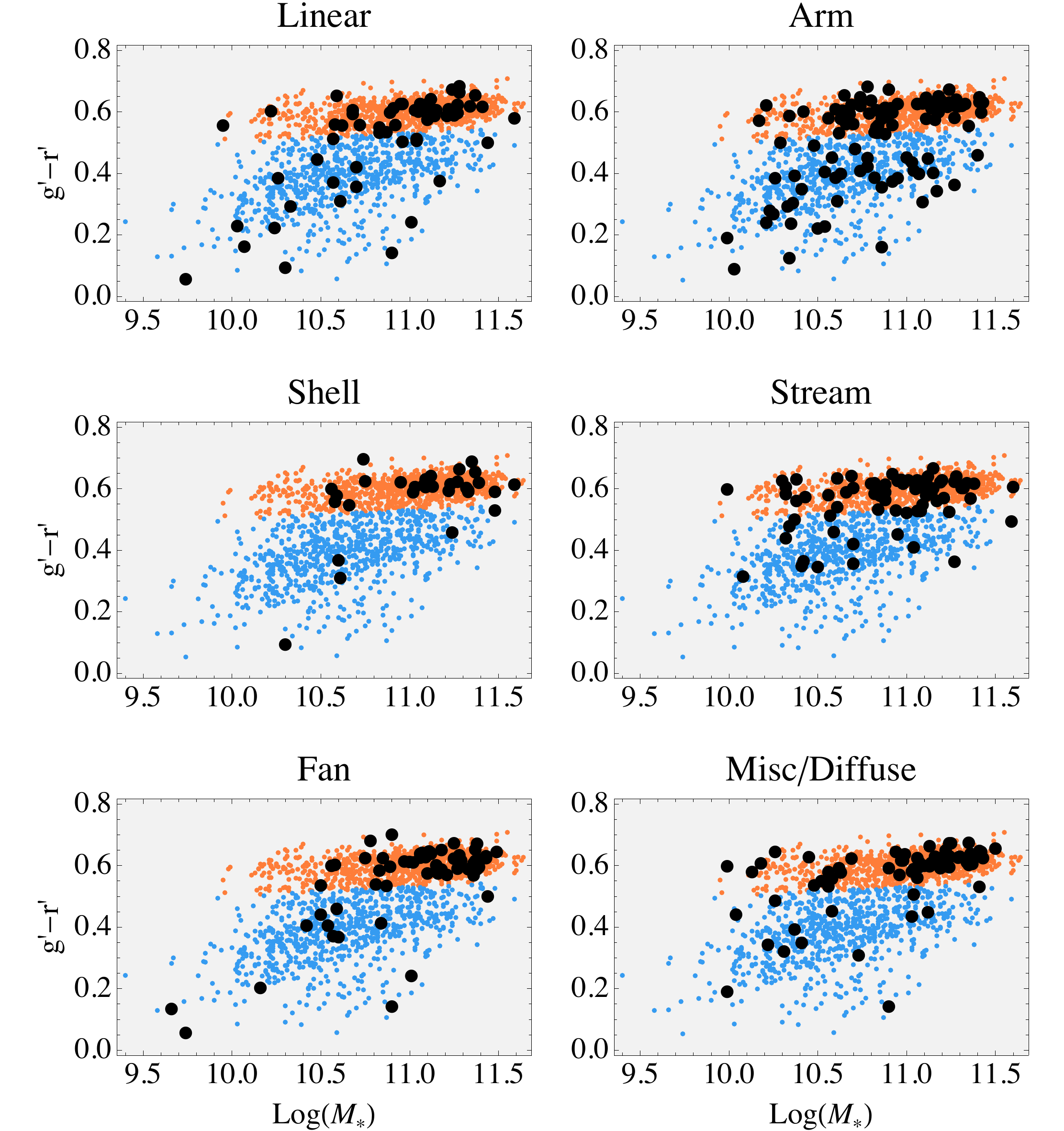}
\caption{Distribution of tidal features as a function of stellar mass and rest-frame $g'-r'$ color. 
Each panel shows the total galaxy population subdivided by color. Plot symbols denote red sequence 
galaxies as red points, and blue cloud galaxies as blue points. In addition, each panel isolates sub-populations 
exhibiting specific tidal features identified by the panel labels. Starting at the top left and running clockwise, the 
panels show the distribution of galaxies with linear structures, arms, streams, miscellaneous/diffuse structures, 
fans, and shells. In each panel the galaxies exhibiting these features are plotted as large black disks. It is clear that 
some of the trends encapsulated by Figures 8 and 9 are strong functions of stellar mass. See text for details.
\label{colormass}}
\end{figure*}

\subsection{Comparison with Previous Work}

Table~1 summarizes the basic results from a number of earlier
investigations into the frequency of tidal structures seen in nearby
galaxies. As we have already noted, there is little consensus.
Setting aside the fact that some surveys span a broad redshift range
over which evolutionary effects may be important (e.g. Bridge et al.
2010), we attribute the lack of homogeneity in the local galaxy
surveys to two main factors: (1) the surveys vary in depth, and (2)
the surveys probe a range of galaxy populations, depending on
selection criteria.

The inhomogeneous depth of the surveys presented in Table~1 has
important implications both in terms of the apparent magnitude limit
of the galaxies probed, and the limiting surface brightness of the
datasets.  The former limitation can, to some extent, be addressed by
our investigation. In the next section we will show how the mix of
tidal features seen in galaxy populations varies with the luminosity
(or stellar mass) of the galaxies being surveyed.  On the other hand,
the latter factor is (to some extent) ameliorated by the fact that it
is presently difficult to undertake any study which probes to a
surface brightness below about 0.5\% of the sky, owing to systematic
effects. The implication is that even very deep investigations are
generally confined to limiting surface brightnesses in a band between
27 mag/arcsec$^2$ and 28 mag/arcsec$^2$ in the V-band. Whether the
abundance of tidal features changes greatly within this band is an
open question, but there are hints that it does. For example, we noted
in the previous section that the fraction of red galaxies in our study
showing evidence for tidal features is a factor of 2--3 lower than
found by \citet{van05} and \citet{tal09}, who investigated the
structure of elliptical galaxies at similar luminosity and redshift to
those studied in the present paper and found that 50-75\% of galaxies
contain such features. This contrasts with the $22.4\pm1.5\%$ of red
galaxies showing high-confidence (class 3 or 4) features in our
investigation. It appears that \citet{van05} probes down to a limiting
surface brightness that is about 0.8 mag/arcsec$^2$ deeper than our
own suggesting that the majority of tidal features in early-type
galaxies are seen at surface brightness near (or below) 28
mag/arcsec$^2$.

Recent work by \citet{kim12} finds significantly fewer instances of
structure (17\%) in Early Type Galaxies in the Spitzer Survey of
Stellar Structure in Galaxies, which is more consistent with the
results of this work. However, it should be noted that a small number
of previously identified features were not detected in galaxies within
their sample, including at least one instance of shells far enough
away from the parent galaxy that they were thought to be outside the
field of view of \citet{kim12}.

Perhaps the most similar work to our own in concept is that of
\citet{mis11} who found that only 6\% of their sample had distinct
features and 19\% showed faint indications of features. Our higher
detection rates may arise in part from differences in the
classification method, as well as the more obvious differences in
depth and seeing. For example, broad features such as fans and shells
are included in our classification system but not theirs. Regardless,
comparison with future studies by this group with larger sample sizes
should prove interesting.

As has been noted, a number of the surveys presented in
Table~\ref{SynthesisOfOtherWork} focus on systems selected in a
particular way. In some cases, selection makes it impossible for fair
comparisons to be made. For example, it is difficult to make a direct
comparison between our work and the sample provided by \citet{mar10}.
These authors find a wealth of substructure in a handful of nearby
galaxies, but many of their targets are well-known systems selected in
advance for prominent substructure and in no way can be considered a
statistical sample. In other cases, more detailed comparisons are
possible.  It is particularly interesting to compare our work with
that of \citet{mal83}, who focused on shells in elliptical galaxies.
We find that $3.2\pm0.5\%$ of the red galaxies in our study have
shells (based on systems with confidence levels 3 and 4), compared to
5.8\% of theirs. It is difficult to directly compare the limiting
depth of CCD-based studies to photographic investigations, but we
suspect that the limiting depths of the studies are comparable, and
that most of the discrepancy stems from the manner in which we have
subdivided tidal features into six categories. It is likely that some
features \citet{mal83} classified as shells would be classified fans
in the present paper (see Figure~\ref{feat} for the general similarity
of the features). Around 8\% of our red galaxy sample exhibit shells
or fans. It is also interesting that \citet{mal83} find shell
frequency increases threefold when samples are restricted to isolated
galaxies. More recent studies seem to disagree on the role of the
environment with \citet{adams12} and \citet{sheen12} finding very
different debris frequencies in rich galaxy clusters.  This may be due
to differing survey depths, however.
 
\subsection{Galaxy Mass-Dependence of Tidal Features}
\label{massdep}

As noted in the previous section, it is interesting to consider
whether the visibility of the tidal features identified in our sample
is dependent on the stellar mass of the tidally disturbed galaxy.
Figure~\ref{colormass} presents the distribution of galaxies with
strong (confidence class 3 and 4) tidal features as a function of
stellar mass and rest-frame $g'-r'$ color. Stellar masses were
estimated from the rest-frame luminosities and colours using the
methodology described in \citet{bel03}, assuming a Kroupa initial mass
function \citep{kro93}. Each panel in Figure~\ref{colormass}
corresponds to a separate tidal feature, and the galaxy population are
colour-coded according to position within the blue cloud (blue points)
or red sequence (red points).  Galaxies exhibiting the tidal features
isolated in each panel are shown as large black points.

Although our sample does not span a large range in stellar mass, a
number of interesting conclusions can be drawn from
Figure~\ref{colormass}. As has already been noted, `linear' features
occur most frequently in red systems, but the figure reveals a strong
mass dependence in the visibility of such features on the red
sequence, with a bias toward the most massive galaxies.  Linear
features (as well as shells and fans; see below) occur much more
frequently in galaxies with stellar masses $>10^{10.5}~{\rm M}_\odot$.

The small number of linear features seen in galaxies in the blue cloud
do not appear to show any mass dependence.  Systems containing
`shells' also show a strong mass dependence, although these features
are sufficiently rare in blue cloud galaxies that we would be unable
to characterize any mass dependence in these objects even if it
existed.  Systems containing `fans' appear to show trends very similar
to those for systems containing `linear' features. Overall, our data
lends support to the claim by van Dokkum (2005) that diffuse fan-like
features might be associated with dry merging activity, since they are
found most frequently in massive red sequence galaxies. Although, it
is worth noting that shells seem an even better tracer of dry mergers,
and that linear features appear to be about as effective as fans for
tracing tidal interactions on the red sequence.

Interestingly, not all tidal features exhibit evidence for a
dependence on stellar mass.  There is little evidence for any mass
dependence in the `arm' classifications in either red or blue
galaxies. The main impression that emerges from Figure~10 for this
class of features is simply a reinforcement of the conclusion from
Figure~9 that tidal arms are the dominant form of tidal feature seen
in systems on the blue cloud.  There also appears to be little
evidence for mass dependence in `streams' in either the red or blue
populations.  If this class of feature traces minor accretion events,
then the implication is that these occur at a similar rate in both red
and blue systems. Finally, there appears to be a hint of bimodality in
the mass dependence of features classified as `miscellaneous/diffuse',
which suggests that, befitting its name, this category contains at
least two independent types of structures.

In an investigation of on-going mergers out to $z=0.7$, \citet{cho11}
show that at blue mergers (with colours typical of the blue cloud) are
far more common than red mergers (with colours typical of the red
sequence), but our findings here for the more local universe indicate
that faint debris is more common in red systems.  If the preponderance
of tidal features in red galaxies is indicative of dry mergers then,
taken together with the \citet{cho11} result, this suggests that
either the accretion rate onto red galaxies has been increasing in the
very recent past or that tidal features in red galaxies are
longer-lasting. It is difficult to draw firm conclusions because
\citet{cho11} explores much further down the faint end of the local
luminosity function than we do.  The presence of tidal features shows
a fairly strong mass dependence, and it is clear from Figure~12 that
the $r'=17$ mag cut strongly biases our sample toward the bright end
of the luminosity function, which is dominated by early-type systems.
An investigation which probes the demographics of tidal structures in
fainter galaxy populations would likely prove interesting.

\section{Conclusions}
Using the wide-field component of the Canada-France-Hawaii Telescope
Legacy Survey, we have analysed the incidence of tidal disturbances in
1781 luminous ($M_{r^\prime}<-19.3$ mag) galaxies in the magnitude
range $15.5 <r^\prime <17$~ mag and in the redshift range $0.04 < z <
0.2$.  We present our results in the form of a catalog that we hope
will form the basis for future investigations into the nature and
origin of low surface brightness debris around galaxies.  Although we
have classified tidal features according to morphology (e.g. streams,
shells and tails), we do not attempt to interpret them in terms of
their physical origin (e.g. post-merger debris versus arising from a
minor accretion event).  We find that around 12\% of the galaxies in
our sample show clear tidal features at the highest confidence level.
This fraction rises to about 18\% if we include systems with
convincing albeit weaker tidal features, and to 26\% if we include
systems with more marginal features that may or may not be tidal in
origin. These proportions are a strong function of rest-frame colour
and of stellar mass.  Linear features, shells, and fans are much more
likely to occur in massive galaxies with stellar masses
$>10^{10.5}~{\rm M}_\odot$, and red galaxies are twice as likely to
show tidal features than are blue galaxies.  We discuss how our
overall statistics compare to those reported in the literature and
conclude that varying photometric depths and selection criteria can
lead to significant variations in the derived debris frequencies.
      
\acknowledgments 

Based on observations obtained with MegaPrime/ MegaCam, a joint
project of CFHT and CEA (DAPNIA), at the Canada-France-Hawaii
Telescope (CFHT) which is operated by the National Research Council
(NRC) of Canada, the Institut National des Science de l'Univers of the
Centre National de la Recherche Scientifique (CNRS) of France, and the
University of Hawaii. This wo is based in part on data products
produced at the Canadian Astronomy Data Centre as part of the
Canada-France-Hawaii Telescope Legacy Survey, a collaborative project
of NRC and CNRS. This research made use of the ``K-corrections
calculator'' service available at http://kcor.sai.msu.ru/. This
research has also made use of SAOImage DS9, developed by Smithsonian
Astrophysical Observatory

\begin{deluxetable}{cccccccccc}
\tabletypesize{\small}
\tablewidth{0pt}
\tablecaption{Master data table\tablenotemark{a} \label{MasterDataTable}}
\tablehead{
\colhead{ID} & 
\colhead{Confidence\tablenotemark{b}} & 
\colhead{Features\tablenotemark{c}} &
\colhead{RA} & 
\colhead{Dec} & 
\colhead{$z_{phot}$} &
\colhead{M$_r'$\tablenotemark{d}} &
\colhead{M$_\star$\tablenotemark{e}} &
\colhead{$(g'-r')$\tablenotemark{f}} &
\colhead{Red/Blue\tablenotemark{g}}
\\
\colhead{} & 
\colhead{} & 
\colhead{} &
\colhead{[J2000]} & 
\colhead{[J2000]} & 
\colhead{} &
\colhead{[mag]} &
\colhead{[$\log({\rm M_\sun}$)]} &
\colhead{[mag]} &
\colhead{}
}
\startdata
  W1-2  &  4  &  L,A,F   &   30.690147  &  -11.165118  &  0.147  &  -22.38  &  11.06  &  0.623  &  R  \\
  W1-3  &  1  &  -       &   30.201271  &  -11.144945  &  0.178  &  -23.43  &  11.24  &  0.413  &  B  \\
  W1-4  &  0  &  -       &   30.391056  &  -11.107487  &  0.182  &  -23.71  &  11.36  &  0.414  &  B  \\
  W1-7  &  4  &  L,S,A   &   30.196264  &  -11.033447  &  0.120  &  -22.10  &  10.95  &  0.623  &  R  \\
  W1-8  &  0  &  -       &   30.493235  &  -11.019429  &  0.113  &  -21.76  &  10.80  &  0.616  &  R  \\
  W1-9  &  0  &  -       &   30.513254  &  -10.806742  &  0.111  &  -21.88  &  10.62  &  0.412  &  B  \\
 W1-11  &  0  &  -       &   31.069776  &  -10.696386  &  0.126  &  -22.02  &  10.67  &  0.402  &  B  \\
 W1-13  &  0  &  -       &   30.596039  &  -10.623502  &  0.152  &  -22.71  &  10.94  &  0.400  &  B  \\
\enddata
\tablenotetext{a}{The full version of this table appears in the electronic version of this paper.}
\tablenotetext{b}{Confidence class. 4=certain, 3=probable, 2= possible, 1=hint, 0=none. See text for details.}
\tablenotetext{c}{Feature classifications. `L'=linear, `A'=arm, `H'=shell, `S'=stream, `F'=fan, `M'=miscellaneous/diffuse.}
\tablenotetext{d}{Rest-frame r'-band magnitude}
\tablenotetext{e}{Base-10 logarithm of stellar mass in solar units.}
\tablenotetext{f}{Rest-frame color.}
\tablenotetext{g}{Flag indicating blue cloud (B) or red sequence (R).}
\end{deluxetable}

\end{document}